\documentclass{iaus}
\usepackage{graphicx}

\title[BBN with long-lived strongly interacting relic particles] 
{Big Bang Nucleosynthesis with long-lived \\ strongly interacting relic particles}

\author[M. Kusakabe, T. Kajino, T. Yoshida \& G. J. Mathews]   
{Motohiko Kusakabe$^{1}$\thanks{Research Fellow of the Japan Society for the Promotion of Science (JSPS).}, Toshitaka Kajino$^{2,3,4}$, Takashi Yoshida$^{2}$ \and Grant J. Mathews$^{5}$}

\affiliation{
$^1$Institute for Cosmic Ray Research, University of Tokyo, Kashiwa, Chiba 277-8582, Japan \\ email: {\tt kusakabe@icrr.u-tokyo.ac.jp} \\[\affilskip]
$^2$Department of Astronomy, Graduate School of Science, University of Tokyo, \\ Bunkyo-ku, Tokyo 113-0033, Japan \\
$^3$Division of Theoretical Astronomy, National Astronomical Observatory
of Japan, \\ Mitaka, Tokyo 181-8588, Japan \\
$^4$Department of Astronomical Science, The Graduate University for
Advanced Studies, \\ Mitaka, Tokyo 181-8588, Japan \\
$^5$Department of Physics, Center for Astrophysics, University of
Notre Dame, \\ Notre Dame, IN 46556, USA}

\pubyear{2010}
\volume{268}  
\jname{Light elements in the Universe}
\editors{C. Charbonnel, M. Tosi, F. Primas \& C. Chiappini, eds.}
\begin{document}

\maketitle

\begin{abstract}
We study effects of relic long-lived strongly interacting massive particles
($X$ particles) on big bang nucleosynthesis (BBN).  The $X$ particle is assumed to have existed
during the BBN epoch, but decayed long before detected.  The interaction strength between an $X$ and a
nucleon is assumed to be similar to that between nucleons.  Rates of nuclear reactions and beta decay of $X$-nuclei are calculated, and the BBN in the presence of neutral charged $X^0$ particles is calculated taking account of captures of $X^0$ by nuclei.  As a result, the $X^0$ particles form bound states with normal nuclei
during a relatively early epoch of BBN leading to the production of
heavy elements.  Constraints on the
abundance of $X^0$ are derived from observations of primordial light element abundances.  Particle models which predict long-lived colored  particles with
lifetimes longer than $\sim$ 200 s are rejected.  This scenario prefers the
production of $^9$Be and $^{10}$B.  There might, therefore, remain a signature of the $X$ particle on
primordial abundances of those elements.  Possible signatures left on light element abundances expected in four different models are summarized.
\end{abstract}

\firstsection 
\section{Introduction}

Primordial lithium abundances are inferred from measurements in metal-poor stars (MPSs).  Observed abundances are roughly constant as a function of metallicity~(\cite[Spite \& Spite 1982]{spi1982}, \cite[Ryan et al. 2000]{rya2000}, \cite[Mel{\'e}ndez \& Ram{\'{\i}}rez 2004]{mel2004}, \cite[Asplund et al. 2006]{asp2006}, \cite[Bonifacio et al. 2007]{bon2007}, \cite[Shi et al. 2007]{shi2007}, \cite[Aoki et al. 2009]{aok2009}) at $^7$Li/H$=(1-2) \times 10^{-10}$.  The theoretical prediction by the standard big bang nucleosynthesis (BBN) model, however, is a factor of $2-4$ higher when its parameter, the baryon-to-photon ratio, is fixed to the value deduced from the observation with Wilkinson Microwave Anisotropy Probe of the cosmic microwave background (CMB) radiation~(\cite[Dunkley et al. 2009]{dun2008}).  A recent study including a reevaluation of nuclear reaction rate of $^3$He($\alpha$,$\gamma$)$^7$Be shows the standard BBN (SBBN) prediction of $^7$Li/H=$(5.24^{+0.71}_{-0.67})\times 10^{-10}$~(\cite[Cyburt et al. 2008]{cyb2008}).  The discrepancy indicates some mechanism of $^7$Li reduction having operated in some epoch from the BBN to this day.  One possible astrophysical process to reduce $^7$Li abundances in stellar surfaces is the combination of the atomic and turbulent diffusion~(\cite[Richard et al. 2005]{ric2005}, \cite[Korn et al. 2006, 2007]{kor200607}).  The precise trend of Li abundance as a function of effective temperature found in the metal-poor globular cluster NGC 6397 is, however, not reproduced yet~(\cite[Lind et al. 2009]{lin2009}).

$^6$Li/$^7$Li isotopic ratios of MPSs have also been measured spectroscopically.  The $^6$Li abundance as high as $^6$Li/H$\sim 6\times10^{-12}$ was suggested~(\cite[Asplund et al. 2006]{asp2006}), which is about 1000 times higher than the SBBN prediction.  Convective motions in the atmospheres of MPSs could cause systematic asymmetries in the observed line profiles and mimic the presence of $^6$Li~(\cite[Cayrel et al. 2007]{cay2007}).  Nevertheless, there still remain a few or several MPSs with certain detections of high $^6$Li abundances after estimations of convection-triggered line asymmetries~(\cite[Garc{\'{\i}}a P{\'e}rez et al. 2009]{gar2009}, \cite[Steffen et al. 2009]{ste2009}).   This high $^6$Li abundance is a problem since the standard Galactic cosmic ray (CR) nucleosynthesis models predict negligible amounts of $^6$Li yields compared to the observed level in the epoch corresponding to the metallicity of [Fe/H] $<-2$~(\cite[Prantzos 2006]{pra2006}).

Be and B abundances are also observed in MPSs.  $^9$Be abundances increase linearly as iron abundance when the Galaxy evolves chemically~(\cite[Boesgaard et al. 1999]{boe1999}, \cite[Primas et al. 2000a]{pri2000a}, \cite[Tan et al. 2009]{tan2009}, \cite[Smiljanic et al. 2009]{smi2009}, \cite[Ito et al. 2009]{ito2009}, \cite[Rich \& Boesgaard 2009]{rich2009}).  In a region of very low metallicities of [Fe/H]$<-3$, a dispersion in Be abundances is indicated~(\cite[Primas et al. 2000b]{pri2000b}, \cite[Boesgaard \& Novicki 2006]{boe2006}).  B abundances increase linearly as iron abundance~(\cite[Duncan et al. 1997]{dun1997}, \cite[Garcia Lopez et al. 1998]{gar1998}, \cite[Primas et al. 1999]{pri1999}, \cite[Cunha et al. 2000]{cun2000}).  So far no primordial plateau abundances of Be and B are found.

As a cosmological solution to the Li problems, BBN models including exotic decaying particles have been studied~(see contribution by Jedamzik, this volume).  Nonthermal nuclear reactions triggered by the radiative decay of long-lived particles can produce $^6$Li nuclides in amounts greater than observed in MPSs and at most $\sim 10$ times as much as the level without causing discrepancies in abundances of other light elements or the CMB energy spectrum~(\cite[Kusakabe et al. 2006, 2009a]{kus2006,kus2009a}).  If negatively-charged leptonic $X^-$ particles exist in the BBN epoch, they affect the nucleosynthesis~(see contribution by Jedamzik, this volume).  The $X^-$ particles get bound to positively charged nuclides with binding energies of $\sim O(0.1-1)$~MeV.  Since the binding energies are low, the bound states between the $X^-$ and nuclides form late in BBN epoch.  Nuclear reactions at this temperature are no longer efficient so that the effect of negatively-charged particles is rather small.  Interestingly the $X^-$ particle can catalyze a preferential production of $^6$Li~(\cite[Pospelov 2007]{pos2007}) and weak destruction of $^7$Be~(\cite[Bird et al. 2008]{bir2007}, \cite[Kusakabe et al. 2007]{kus2007}).  Non-equilibrium nuclear network calculation of this model~(\cite[Kusakabe et al. 2008]{kus2008a}) with realistic cross sections derived from a quantum mechanical calculation~(\cite[Kamimura et al. 2009]{kam2009}) shows that $^6$Li production and $^7$Li reduction can simultaneously occur and that there are no likely signature in primordial abundances of Be and heavier nuclei.

As an astrophysical solution to the $^6$Li problem, the cosmological CR nucleosynthesis associated with a possible activity of supernova explosions in the early epoch of structure formation has been suggested~(\cite[Rollinde et al. 2005, 2006]{rol200506}).  This $^6$Li production mechanism is likely realized with coproduction of $^9$Be and $^{10,11}$B nuclides in abundances probably within reach of future observations of MPSs~(\cite[Kusakabe 2008]{kus2008b}, \cite[Rollinde et al. 2008]{rol2008}).  The $^6$Li production is, however, not a certain possibility since a calculation in a similar scenario but using a different star formation and chemical evolution history fails to reproduce the $^6$Li plateau level~(\cite[Evoli et al. 2008]{evo2008}).

In this paper we show a new BBN scenario which includes long-lived strongly interacting relic particles which appear in some particle models beyond the standard.  Constraints on their abundance and lifetime is derived.  Signatures of such relic particles are found to be possibly left on the primordial abundances of Be and B.  A theoretical prediction of primordial light element abundances are given in the limit of long lifetime.

\section{Model}

\begin{figure}[t]
\begin{center}
 \includegraphics[width=3.4in]{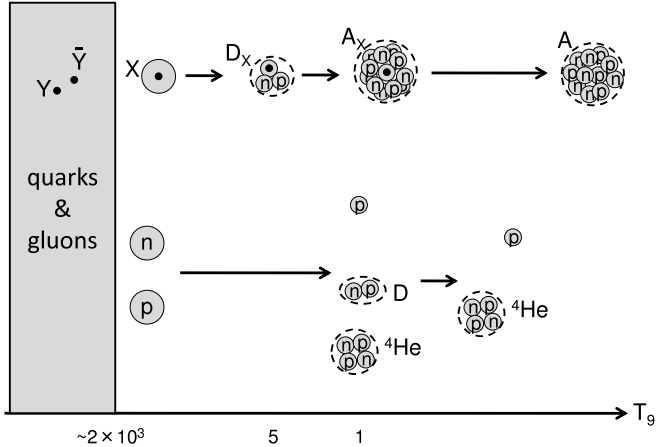} 
 \caption{Nucleosynthesis in the presence of the long-lived strongly interacting $X$ particle.}
   \label{fig1}
\end{center}
\end{figure}

Some particle models include long-lived heavy colored particles, denoted by $Y$~(\cite[Kang et al. 2008]{kan2008}).  If the hypothetical colored particle $Y$ exists, they would have experienced production and annihilation in the early universe.  When the temperature of the universe decreases to $T < 180$~MeV, $Y$ gets confined in exotic hadrons, which we call $X$ (see Fig.\ref{fig1}).  Two X particles can form bound states and the bound states decay into lower energy states gradually.  $Y$ particles in $X$ particles could annihilate eventually when two $X$ particles approach.  Taking account of this process, \cite{kan2008} estimated the relic abundance of $X$ to be $n_X\sim 10^{-8} n_b$, where $n_X$ and $n_b$ are number densities of $X$ and baryon, respectively.  If the strongly interacting $X$ particle had existed in BBN epoch, it would have affected the nucleosynthesis.  We study effects of $X$ on BBN.  The $X$ particle is assumed to be of spin $0$, charge $0$ and mass much larger than nuclear mass of $O$(1~GeV).  The strength of interaction between an $X^0$ particle and nuclei is supposed to be similar to that between a nucleon and the nuclei as a rough approximation.

Binding energies between nuclides and an $X^0$ are calculated.  The adopted nuclear potential between a nucleon and an $X$ is well type reproducing the binding energy of $n$+$p$ system when used for the $n$+$p$ system.  The Woods-Saxon potential is adopted for systems of nuclides and an $X$.  Two-body Shr\"{o}dinger equations are solved by a variational calculation, and binding energies are derived.  The binding energies are of $O$(10~MeV), and this strong binding leads to bound state formations early in the BBN epoch.

Thermonuclear reaction rates and $\beta$-decay rates of $X$-bound nuclei (i.e., $X$-nuclei or $A_X$) are, then, calculated.  Reaction $Q$-values are derived taking account of binding energies of $X$-nuclei.  The rates associated with $X$-nuclei are estimated using available cross sections for normal nuclei by correcting for $Q$-values and mass numbers of reactants.

We have put radiative $X$ capture reactions by nuclides, nuclear reactions and $\beta$-decay of $X$-nuclides and all inverse reactions into the SBBN network code~(\cite[Kawano 1992]{kawano}) and solved a set of rate equations.  See~\cite{kus2009b} for details on this study.

\section{Result}

Figure~\ref{fig1} shows a schematic view of nucleosynthesis as a function of $T_9\equiv T/(10^9$~K$)$ with temperature $T$.  The relic abundance of $X$ would be rather small ($Y_X=n_X/n_b\sim 10^{-8}$) although it depends on the mass and interaction strength of the exotic colored $Y$ particle.  The main story of BBN, therefore, does not change even if the strongly interacting $X$ particle exists.  Effects on abundant light elements, i.e., H and He are small, while those on minor elements, i.e., Li, Be and B are significant.  At high temperature of $T_9 > 5$, the $X$ particles exist in the free state.  When the temperature decreases to $T_9 \sim 5$, $X$ particles radiatively capture neutrons and protons to form bound states.  $n_X$ and $H_X$ thus formed are processed to $^2$H$_X$ and $^3$He$_X$.  At $T_9\sim 5$, the normal D abundance temporarily increases during the $^4$He production.  An efficient nuclear process of $X$-nuclei then occurs through consecutive nonradiative strong D-capture reactions, i.e., ($d$,$p$) and ($d$,$n$).  Finally the $X$ particles decay and heavy nuclides are left.  From the fact that all $X$ particles form $X$-nuclei before their decay, it can be seen that strongly interacting exotic particles have a great impact on BBN~(\cite[Kusakabe et al. 2009b]{kus2009b}).  One reason is that binding energies between an $X$ and nuclides are large, and cross sections of captures of nucleons by $X$ are large, leading to early formation of bound states between nucleons and an $X$.  Another reason is that $^5$He, $^5$Li and $^8$Be which are unstable to particle decays can be stabilized against the decays when they are bound to $X$ particles, so that heavy $X$-nuclides can form through those stabilized states.

\begin{figure}[t]
\begin{center}
 \includegraphics[width=3.4in]{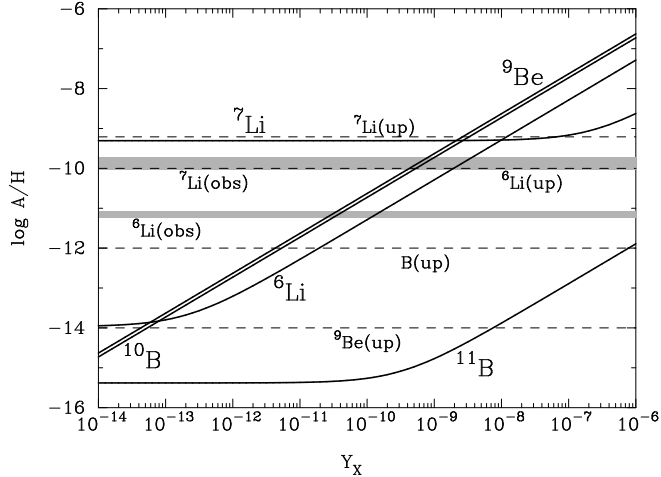} 
  \caption{Calculated abundances of $^{6,7}$Li, $^9$Be and $^{10,11}$B (solid lines) as a function of the $X$ abundance $Y_X$.  The grey bands correspond to the observed mean values of $^7$Li abundances and $^6$Li abundances of MPSs with detection of the isotope.  The dashed lines indicate the adopted upper limits on the primordial abundances (see text).}
   \label{fig2}
\end{center}
\end{figure}

Figure~\ref{fig2} shows calculated abundances of $^{6,7}$Li, $^9$Be and $^{10,11}$B (solid lines) as a function of the initial $X$ abundance $Y_X$ in the case of very long lifetime, $\tau_X$, compared to the BBN time scale.  Resulting yields through the $X$-catalyzed nucleosynthesis are all proportional to $Y_X$.  The prediction of primordial abundances in this scenario is as follows:
\begin{eqnarray}
^6{\rm Li/H} \sim ^9{\rm Be/H}\sim ^{10}{\rm B/H} \sim 10^{-9}\left(Y_X/10^{-8}\right).\\
\label{eq1}
^7{\rm Li/H} \sim 10^{-11}\left(Y_X/10^{-8}\right).\\
\label{eq2}
^{11}{\rm B/H} \sim 10^{-14}\left(Y_X/10^{-8}\right).
\label{eq3}
\end{eqnarray}
The grey bands correspond to the observed mean values of $^7$Li abundances~(\cite[Ryan et al. 2000]{rya2000}) and $^6$Li abundances of MPSs with detection of the isotope~(\cite[Asplund et al. 2006]{asp2006}).  The dashed lines show the adopted upper limits on the primordial abundances from observations:  $^7$Li/H$<5\times 1.23\times 10^{-10}$ allowing for large depletion factors of up to five, $^6$Li/H$<5\times 10^{-10}$ allowing for depletion factors up to $\sim 10$, $^9$Be/H$<10^{-14}$~(\cite[Ito et al. 2009]{ito2009}) and B/H$< 10^{-12}$~(\cite[Garcia Lopez et al. 1998]{gar1998}).  $Y_X$ values leading to overproductions of light nuclides in abundances above the observational upper limits should be ruled out.  The upper limits on the primordial Be abundance is found to give the most stringent constraint on $X$ abundance, i.e., $Y_X< 10^{-13}$ in the case of long lifetime.

When the lifetime $\tau_X$ is not much longer than the BBN time scale, the decay of $X$ should be considered in the numerical calculation.  A constraint on the initial abundance $Y_X$ and the lifetime $\tau_X$ is derived~(\cite[Kusakabe et al. 2009b]{kus2009b}).  A solution to the $^6$Li or $^7$Li problems was not found in this paradigm.  As seen in Fig.\ref{fig2}, the light element abundances monotonically increase by the existence of $X$.  A reduction of $^7$Li abundance is thus NOT possible.  The $^6$Li production is realized with overproduction of $^9$Be and $^{10}$B, which is excluded from observations of Be and B. A $^6$Li production is then NOT possible.

Constraints on the $Y_X$ and $\tau_X$ are derived from limits on Li, Be, B abundances depending on lifetime ranges.  Two important predictions of this scenario is found.  First, $^9$Be and B can be produced in amounts more than SBBN predictions.  Future observations of Be and B abundances in MPSs might show plateaus suggesting a primordial origin.  Second, $^{10}$B tends to be produced much more than $^{11}$B, i.e., a resulting isotopic ratio $^{10}$B/$^{11}$B is very high (see Fig.\ref{fig2}).  This preferential production of $^{10}$B has never been predicted in other models of Be production like the Galactic or cosmological CR nucleosynthesis ($^{10}$B/$^{11}$B$\sim 0.4$: e.g.,~\cite[Ramaty et al. 1997]{ram1997}, \cite[Kusakabe 2008]{kus2008b}) or the supernova neutrino process ($^{10}$B/$^{11}$B$\ll 1$: e.g.,~\cite[Woosley \& Weaver 1995]{woo1995}, \cite[Yoshida et al. 2005]{yos2005}).

A constraint on $\tau_X$ is derived from this result and estimations of initial abundance of $X$.  If $Y_X\sim 10^{-8}$~(\cite[Kang et al. 2008]{kan2008}) is adopted, its lifetime should be $\tau_X < 200$~s.

\section{Summary}

The BBN scenario in the presence of long-lived strongly interacting relic particle $X$ is described.  Particles which interact with nuclei as a nucleon does capture nucleons to form $X$-nuclei in an early stage of BBN.  The $X$-nuclei capture deuterons and increase their mass numbers.  This nuclear catalysis of $X$ activates new reaction paths to heavy nuclei.  Constraints on the lifetime and abundance of $X$ are derived.  The lifetime should be less than 200~s.

\begin{table}[t]
  \begin{center}
  \caption{Possibility to solve the Li problems and expected signatures in elemental abundances.}
  \label{tab1}
 {\scriptsize
  \begin{tabular}{|l|c|c||c|}\hline 
Model & $^6$Li Problem solved ? & $^7$Li Problem solved ? & Possible Signatures on Other Nuclides\\ \hline
radiative decay & {\bf YES} & NO & NO \\ \hline
leptonic $X^-$ & {\bf YES} & {\bf YES} & {\bf NO} \\ \hline
strongly interacting $X^0$ & {\bf NO} & {\bf NO} & {\bf $^9$Be and/or $^{10}$B} \\ \hline
early cosmic ray & {\bf YES} & NO & {\bf $^9$Be and $^{10, 11}$B} \\ \hline
  \end{tabular}
  }
 \end{center}
\vspace{1mm}
\end{table}

Table \ref{tab1} summarizes possibilities for four models to be solutions to the $^6$Li or $^7$Li problems and expected signatures in abundances of other light nuclides.  In light of all available observational data, nonstandard BBN processes (first two models) possibly operate to produce $^6$Li and/or reduce $^7$Li.  The existence of long-lived relic particles should be ascertained with future astronomical observations and collider experiments.  The cosmological CR nucleosynthesis might also have produced light elements in an early epoch of the structure formation.  Observations of primordial abundances of Li, Be and B are very important to constrain the possible processes in the early universe.

\begin{acknowledgments}
This work was supported by the Mitsubishi Foundation, MEXT KAKENHI
 (20105004), JSPS KAKENHI (20244035), JSPS Core-to-Core Program EFES, and Grant-in-Aid for JSPS Fellows (21.6817).  Work at the University of Notre Dame
 was supported by the U.S. Department of Energy under Nuclear Theory Grant DE-FG02-95-ER40934.
\end{acknowledgments}


\begin{thebibliography}{}

\bibitem[Aoki et al.(2009)]{aok2009} {Aoki, W., et al.}\ 2009, \textit{ApJ}, 698, 1803 

\bibitem[Asplund et al.(2006)]{asp2006} {Asplund, M., et al.}\ 2006, \textit{ApJ}, 644, 229 

\bibitem[Bird et al.(2008)]{bir2008} {Bird, C., Koopmans, K., 
\& Pospelov, M.}\ 2008, \textit{Phys. Rev. D}, 78, 083010 

\bibitem[Boesgaard et al.(1999)]{boe1999} {Boesgaard, A.~M., et al.}\ 1999, \textit{AJ}, 117, 1549 

\bibitem[Boesgaard \& Novicki(2006)]{boe2006} {Boesgaard, A.~M., \& Novicki, M.~C.}\ 2006, \textit{ApJ}, 641, 1122 

\bibitem[Bonifacio et al.(2007)]{bon2007} {Bonifacio, P., et al.}\ 2007, \textit{A\&A}, 462, 851 

\bibitem[Cayrel et al.(2007)]{cay2007} {Cayrel, R., et al.}\ 2007, \textit{A\&A}, 473, L37 

\bibitem[Cunha et al.(2000)]{cun2000} {Cunha, K., Smith, V.~V., 
Boesgaard, A.~M., Lambert, D.~L.}\ 2000, \textit{ApJ}, 530, 939 

\bibitem[Cyburt et al.(2008)]{cyb2008} {Cyburt, R.~H., Fields, 
B.~D., 
\& Olive, K.~A.}\ 2008, \textit{J. Cosmol. Astropart. Phys.}, 11, 12 

\bibitem[Duncan et al.(1997)]{dun1997} {Duncan, D.~K., et al.}\ 1997, \textit{ApJ}, 488, 338 

\bibitem[Dunkley et al.(2009)]{dun2009} {Dunkley, J., et al.}\ 
2009, \textit{ApJS}, 180, 306 

\bibitem[Evoli et al.(2008)]{evo2008MNRAS} {Evoli, C., Salvadori, S., 
\& Ferrara, A.}\ 2008, \textit{MNRAS}, 390, L14 

\bibitem[Garcia Lopez et al.(1998)]{gar1998} {Garcia Lopez, R.~J., et al.}\ 1998, \textit{ApJ}, 500, 241 

\bibitem[Garc{\'{\i}}a P{\'e}rez et al.(2009)]{gar2009} {Garc{\'{\i}}a P{\'e}rez, A.~E., et al.}\ 2009, \textit{A\&A}, 504, 213 


\bibitem[Ito et al.(2009)]{ito2009} {Ito, H., Aoki, W., Honda, 
S., Beers, T.~C.}\ 2009, \textit{ApJ} (Letters), 698, L37 

\bibitem[Kamimura et al.(2009)]{kam2009} {Kamimura, M., Kino, 
Y., \& Hiyama, E.}\ 2009, \textit{Progress of Theoretical Physics}, 121, 1059 

\bibitem[Kang et al.(2008)]{kan2008} {Kang, J., Luty, M.~A., 
\& Nasri, S.}\ 2008, \textit{J. High Energy Phys.}, 9, 86 

\bibitem[Kawano(1992)]{kawano} {Kawano, L.}\ 1992, \textit{NASA STI/Recon Technical Report N}, 92, 25163 

\bibitem[Korn et al.(2006)]{kor200607} {Korn, A.~J., et al.}\ 2006, \textit{Nature}, 442, 657;
\ 2007, \textit{ApJ}, 671, 402 

\bibitem[Kusakabe et al.(2006)]{kus2006} {Kusakabe, M., Kajino, 
T., \& Mathews, G.~J.}\ 2006, \textit{Phys. Rev. D}, 74, 023526 

\bibitem[Kusakabe et al.(2007)]{kus2007} {Kusakabe, M., et al.}\ 2007, \textit{Phys. Rev. D}, 76, 121302
\bibitem[Kusakabe et al.(2008)]{kus2008a} {Kusakabe, M., et al.}\ 2008, \textit{ApJ}, 680, 846 

\bibitem[Kusakabe (2008)]{kus2008b} {Kusakabe, M.}\ 2008, \textit{ApJ}, 
681, 18 


\bibitem[Kusakabe et al.(2009a)]{kus2009a} {Kusakabe, M., et al.}\ 2009a, \textit{Phys. Rev. D}, 79, 123513 

\bibitem[Kusakabe et al.(2009b)]{kus2009b} {Kusakabe, M., Kajino, 
T., Yoshida, T., Mathews, G.~J.}\ 2009b, \textit{Phys. Rev. D}, 80, 103501 

\bibitem[Lind et al.(2009)]{lin2009} {Lind, K., et al.}\ 2009, \textit{A\&A}, 503, 545 

\bibitem[Mel{\'e}ndez \& Ram{\'{\i}}rez(2004)]{mel2004} {Mel{\'e}ndez, J., \& Ram{\'{\i}}rez, I.}\ 2004, \textit{ApJ} (Letters), 615, L33 

\bibitem[Pospelov(2007)]{pos2007} {Pospelov, M.}\ 2007, \textit{Phys. Rev. Lett.}, 98, 231301 

\bibitem[Prantzos(2006)]{pra2006} {Prantzos, N.}\ 2006, \textit{A\&A}, 448, 665 

\bibitem[Primas et al.(1999)]{pri1999} {Primas, F., Duncan, D.~K., Peterson, R.~C., Thorburn, J.~A.}\ 1999, \textit{A\&A}, 343, 545

\bibitem[Primas et al.(2000a)]{pri2000a} {Primas, F., Molaro, P., Bonifacio, P., Hill, V.}\ 2000, \textit{A\&A}, 362, 666 

\bibitem[Primas et al.(2000b)]{pri2000b} {Primas, F., Asplund, M., Nissen, P.~E., Hill, V.}\ 2000, \textit{A\&A}, 364, L42 

\bibitem[Ramaty et al.(1997)]{ram1997} {Ramaty, R., Kozlovsky, B., Lingenfelter, R.~E., Reeves, H.}\ 1997, \textit{ApJ}, 488, 730 

\bibitem[Rich \& Boesgaard(2009)]{ric2009} {Rich, J.~A., \& Boesgaard, A.~M.}\ 2009, \textit{ApJ}, 701, 1519 

\bibitem[Richard et al.(2005)]{ric2005} {Richard, O., Michaud, 
G., \& Richer, J.}\ 2005, \textit{ApJ}, 619, 538 

\bibitem[Rollinde et al.(2005)]{rol200506} {Rollinde, E., 
Vangioni, E., \& Olive, K.}\ 2005, \textit{ApJ}, 627, 666;
\ 2006, \textit{ApJ}, 651, 658 

\bibitem[Rollinde et al.(2008)]{rol2008} {Rollinde, E., et al.}\ 2008, \textit{ApJ}, 673, 676 

\bibitem[Ryan et al.(2000)]{rya2000}
 {Ryan, S.~G., et al.}\ 2000, \textit{ApJ} (Letters), 530, L57 

\bibitem[Shi et al.(2007)]{shi2007} {Shi, J.~R., et al.}\ 2007, \textit{A\&A}, 465, 587 

\bibitem[Smiljanic et al.(2009)]{smi2009} {Smiljanic, R., et al.}\ 2009, \textit{A\&A}, 499, 103 



\bibitem[Spite \& Spite(1982)]{spi1982}
{Spite, F., \& Spite, M.}\ 1982, \textit{A\&A}, 115, 357 

\bibitem[Steffen et al.(2009)]{ste2009} {Steffen, M., et al.}\ 2009, arXiv:0910.5917 

\bibitem[Tan et al.(2009)]{tan2009} {Tan, K.~F., Shi, J.~R., 
\& Zhao, G.}\ 2009, \textit{MNRAS}, 392, 205 

\bibitem[Woosley \& Weaver(1995)]{woo1995} {Woosley, S.~E., \& Weaver, T.~A.}\ 1995, \textit{ApJS}, 101, 181 

\bibitem[Yoshida et al.(2005)]{yos2005}
 {Yoshida, T., Kajino, T., \& Hartmann, D.~H.}\ 2005, \textit{Phys. Rev. Lett.}, 94, 231101 

\end{thebibliography}
\end{document}